\documentstyle[12pt]{article}
  \advance\oddsidemargin  by -1in
  \advance\evensidemargin by -1in
\def\lsim{\mathrel{\rlap{\lower4pt\hbox{\hskip1pt$\sim$}}
    \raise1pt\hbox{$<$}}}         
\def\gsim{\mathrel{\rlap{\lower4pt\hbox{\hskip1pt$\sim$}}
    \raise1pt\hbox{$>$}}}         

\def\be{\begin{equation}}
\def\ee{\end{equation}}
\def\bq{\begin{eqnarray}}
\def\eq{\end{eqnarray}}
\def\bm{\boldmath}

\begin{document}
\pagestyle{empty}

\hfill{\large DFTT 24/96}
 
\hfill{\large May 1996}
 
\vspace{2.0cm}
 
\begin{center}
 
{\large \bm \bf  A CONFINEMENT MODEL CALCULATION OF 
$h_1(x)$ \\}

\vspace{1.0cm}

{\large V.~Barone$^{a,}$\footnote{Also at II
Facolt{\`a} di Scienze MFN, 15100 Alessandria, Italy.},
T.~Calarco$^{b}$ and A.~Drago$^{b}$ \\}

\vspace{1.0cm} 

{\it $^{a}$Dipartimento di
Fisica Teorica, Universit\`a di Torino\\
and INFN, Sezione di
Torino,    10125 Torino, Italy \medskip\\
$^{b}$Dipartimento di Fisica, Universit{\`a} di Ferrara \\
and INFN, Sezione di Ferrara, 44100 Ferrara, Italy \medskip\\ }

\vspace{1.0cm}

{\large \bf Abstract \bigskip\\ }

\end{center}
 
The transverse polarization distribution of quarks $h_1(x)$ 
is computed in a confinement model, the chiral chromodielectric
model. The flavor structure of $h_1$, its $Q^2$ evolution 
and Soffer's inequality 
are studied. The Drell--Yan double 
transverse asymmetry $A_{TT}$
is evaluated and found to be one order of magnitude
smaller than the double longitudinal asymmetry.

\vfill
 
\pagebreak

\baselineskip 20 pt
\pagestyle{plain}

The interest in the transverse polarization distribution of quarks
and antiquarks,  customarily called $h_1(x)$, has been recently strengthened
by the perspective of its possible measurement in future collider 
experiments. Originally introduced by Ralston and Soper \cite{RS}, 
who called
it $h_T(x)$, $h_1(x)$ has been studied in detail, 
from a formal point of view, in more recent papers \cite{JJ,AM}. 
The possible ways
of measuring $h_1(x)$ in various hard processes have been also 
thouroughly investigated \cite{AM,CPR,Ji,JJ2}. Despite this intensive 
work, 
not much is actually known about the shape and the magnitude of
$h_1(x)$. This is of course not surprising since $h_1$, like
all quark distributions, cannot be derived 
from the fundamental theory of strong interactions, QCD. 
At present we possess only: {\it i)} an 
admittedly crude evaluation of $h_1(x)$ in the simplest version 
of the MIT bag model \cite{JJ}, and {\it ii)} 
an estimate of its first moment -- the 
so--called tensor charge -- obtained by QCD sum rule methods \cite{HJ}. 
It is clear that a more sophisticated model calculation 
of $h_1(x)$ is called for. This would also provide useful
indications about the concrete possibility of a measurement of $h_1(x)$
in the experiments which are now being planned. 

The difficulty of an experimental determination of $h_1(x)$,
which is a leading twist quantity, resides mainly in the fact
that, being a chirally-odd distribution, it is not measurable
in polarized deep inelastic scattering. The best method 
to extract $h_1(x)$ seems to be \cite{CPR,Ji} the Drell--Yan dilepton 
production with two transversely polarized proton beams, 
an experiment which will be performed in the near future at 
RHIC \cite{RHIC}.
The Drell--Yan double transverse asymmetry $A_{TT}$ contains information
on the flavor structure of $h_1$. Therefore it would be important 
to predict the magnitude of $A_{TT}$ in order to test the 
feasibility of the experiment. The available model calculations 
of $A_{TT}$ \cite{Ji,BS} 
are all based on the assumption that $h_1^q$ is approximately
equal to the helicity distribution $\Delta q$. Although this is probably
true at very low momentum scales, such as those at which 
confinement model computations are implicitly performed ($Q^2 
\lsim 1 $ GeV$^2$, see below), 
it is certainly not true at experimental $Q^2$ scales (say $Q^2 
\ge 10 $ GeV$^2$). The reason is that $\Delta q$ and $h_1^q$
evolve differently in $Q^2$. Whereas the first moment
of $\Delta q$ is constant,
the first moment of $h_1^q$ decreases
	with increasing $Q^2$, and  
 the evolution in the $x$-shape is even more dramatically different.
Again, a more firmly based evaluation 
of $A_{TT}$ is needed to check whether this quantity is comparable 
in magnitude with the double longitudinal asymmetry $A_{LL}$, 
as it is claimed in \cite{Ji,BS}. 

Another issue which is certainly worth exploring in the framework 
of confinement models is the 
inequality among leading twist polarized and unpolarized 
distribution functions recently derived by 
Soffer \cite{Soffer} (see also \cite{GJJ}) and the possibility of 
its saturation. 

In the following we shall provide a theoretical determination of 
$h_1(x)$ in a confinement model, the chiral chromodielectric
model \cite{Pirner}, 
which has been already succesfully used to compute other 
leading twist structure functions and various nucleon          
properties. In particular, the flavor structure of $h_1$
will be described in detail and the effects of the peculiar 
QCD evolution of $h_1$ investigated. 
A prediction for $A_{TT}$ will also be presented. As we shall see, 
due to the different evolution of $h_1^{q}$ and $\Delta q$, 
$A_{TT}$ turns out to be much smaller than $A_{LL}$.

The quark transverse polarization distribution  reads \cite{JJ}
\be
h_1(x)=\frac{\sqrt{2}} {4\pi}\int{\rm d}\xi^-
e^{-i x p^+\xi^-}
\langle NS_\perp|\psi_+ ^\dagger(\xi)
\gamma_\perp\gamma_5 \psi_+ (0)| NS_\perp\rangle
\vert_{\xi^+=\xi_\perp=0}.
\label{defh1}
\ee
A similar expression holds for the antiquark distribution, with the 
exchange of $\psi$ and $\psi^\dagger$. Note that in eq.~(\ref{defh1})
 only the `good'
light-cone components of the fields, $\psi_+={1\over 2}\gamma_-\gamma_+\psi$, 
appear, signaling
that $h_1$ is a leading twist quantity. 
The $h_1$ distribution
measures the difference in the number of quarks with
transverse polarization  parallel ($\uparrow$) 
and anti-parallel ($\downarrow$) to the proton transverse polarization.
This can be made transparent by introducing 
the Pauli--Lubanski projectors $P_\perp^{\uparrow \downarrow}
=\frac{1}{2} (1 \pm \gamma_{\perp} \gamma_5)$ and inserting a complete
set of states $\{ \vert X \rangle \}$ in eq.~(\ref{defh1}). We then get
\be
h_1(x) = \frac{1}{\sqrt{2}} \, 
\sum_X \{ \vert\langle NS_\perp|P_\perp^{\uparrow} 
\psi_+(0)\vert X \rangle\vert^2-
\vert\langle NS_\perp|P_\perp^{\downarrow}\psi_+(0)\vert X \rangle\vert^2 
\} \, \delta[(1-x)p^+ - p_X^+] \,.
\label{2}
\ee

In a (projected) mean-field approximation,
the matrix elements in eq.~(\ref{2}) can be rewritten
in terms of single--particle (quark or antiquark) matrix elements. For 
a flavor $f$ one thus gets
\bq
h_1^f(x)={1\over\sqrt{2}}
\sum_\alpha\sum_m P(f,\alpha,m) \int  
\frac{{\rm d}\mbox{\bm $p$}_\alpha}{(2\pi)^3(2 p_\alpha^0)}\,
A_\alpha(p_\alpha) \, \delta[(1-x)p^+ - p^+_\alpha]\,\nonumber\\
\;\;\;\;\;\;\;\;\;\; \;\;\;\;\;\; \times \; \overline\varphi(p_\alpha,m)
\gamma_+\gamma_\perp\gamma_5\varphi(p_\alpha,m)
\, ,
\label{ingredients}
\eq
where $\varphi$ is the single-quark wave function, $m$
is the projection of the quark spin  along the direction of the nucleon's
spin, $P(f, \alpha, m)$ is the probability of extracting a quark of
flavor $f$ and spin $m$ leaving a state generically labelled by the 
quantum number $\alpha$.
The overlap function $A_\alpha(p_\alpha)$ 
contains the details of the intermediate states and 
of the projection used to obtain a
nucleon with definite linear momentum 
from a three--quark bag (see for instance \cite{Thomas,noi}).
The intermediate states which contribute to 
eqs.~(\ref{2},\ref{ingredients}) are $2q$ and $3q 1\bar q$ states
for the quark distribution, and $4q$ states for the antiquark distribution.

At this point, we can already 
discuss qualitatively the Soffer inequality \cite{Soffer}, which reads
\be
q^f(x)+\Delta q^f(x) \ge 2 \vert h_1^f(x) \vert \,,
\label{eq4}
\ee
 where $\Delta q^f$ is
the helicity distribution function and $q^f$ the unpolarized 
density. This relation has been proved
in the parton model \cite{Soffer,GJJ} (for a QCD--improved 
parton model discussion of Soffer's inequality see \cite{B}), 
and is satisfied  flavor by flavor by
both the quark and the antiquark distributions. 
An interesting issue is whether Soffer's inequality
is saturated in some quark model (which means that 
$\vert h_1^q \vert$ takes its maximal value). To clarify this problem
let us write the various leading twist distributions
in an explicit form
\begin{eqnarray}
q^f(x) &=&
\sum_{\alpha} \sum_m
 P(f,\alpha,m)
            \,F_\alpha(x)\\ 
\label{eq5}
\Delta q^f(x)&=&
\sum_\alpha \sum_m P(f,\alpha,m) \, (-1)^{(m+3/2)}
            \,G_\alpha(x)\\
\label{eq6}
h_1^f(x) &=&
\sum_\alpha \sum_m P(f,\alpha,m) \, (-1)^{(m+3/2)} 
            \,H_\alpha(x)\,,
\label{eq7}
\end{eqnarray}
where
\begin{eqnarray}
\left. \begin{array}{c}
F_\alpha(x)\\G_\alpha(x)\\H_\alpha(x) \end{array}\right\}
&=& \int\frac{{\rm d}\mbox{\bm $p$}_\alpha}{(2\pi)^3(2 p_\alpha^0)}
A_\alpha(p_\alpha)\delta[(1-x)p^+ - p^+_\alpha]\nonumber\\
&\times &\frac{1}{2} \left\{ \begin{array}{c}
u^2(p_\alpha)
+2u(p_\alpha)v(p_\alpha)\frac{p_\alpha^z}{\vert \mbox{\bm $p$}_\alpha \vert}
+v^2(p_\alpha)\\
u^2(p_\alpha)
+2u(p_\alpha)v(p_\alpha)\frac{p_\alpha^z}{ \vert \mbox{\bm $p$}_\alpha \vert}
+v^2(p_\alpha)(2 \left 
(\frac{p_\alpha^z}{\vert \mbox{\bm $p$}_\alpha \vert}\right )^2-1) \\
u^2(p_\alpha)
+2u(p_\alpha)v(p_\alpha)\frac{p_\alpha^z}{\vert \mbox{\bm $p$}_\alpha \vert}
+v^2(p_\alpha)(1- 
\left (\frac{p_\alpha^\perp}{\vert \mbox{\bm $p$}_\alpha \vert}\right )^2)
\end{array}\right. \, .   
\label{correnti}
\end{eqnarray}
In eq.~(\ref{correnti}) $p^{\perp}$ is the projection of the momentum 
in the plane perpendicular to the proton's trajectory (chosen to be 
the $z$ axis), 
and the currents have been
written in terms of the single quark wave-function
in momentum space
\begin{equation}
\varphi (p,m)=\left(
\begin{array}{c} u(p)\\ \mbox{\bm $\sigma\cdot \hat p$} 
\, v(p)\end{array}\right )
\,\chi_m \, .
\label{eq9}
\end{equation}

Notice that the three quantities $F_{\alpha}, G_{\alpha}, H_{\alpha}$
satisfy the equality: $F_{\alpha}(x)
+ G_{\alpha}(x) = 2 \, H_{\alpha}(x)$. This has led to the erroneous 
conclusion \cite{Soffer} that the inequality (\ref{eq4})
is saturated for a relativistic quark model, such as the MIT bag model.
It is clear from eqs.~(5-7) that the spin--flavor structure
of the proton, which results in the appearance of 
the probabilities $P(f,\alpha,m)$, spoils this argument
and prevents in general the saturation of the inequality. 

Soffer's inequality is saturated only in very specific (and
somehow unrealistic) cases.
For instance, it is saturated when $P(f,\alpha,-1/2)=0$, 
which happens if the proton is modeled as a bound state
of a scalar diquark and a $u$ quark. Of course, this is 
too a rough picture of the proton. However, it is interesting 
to note that 
in $SU(6)$ the  $\Lambda$ is a bound state of a scalar--isoscalar
$ud$ diquark and an $s$ quark: the $h_1$ distribution of the latter 
then attains the maximal value compatible with (\ref{eq4}).

Another instance of saturation 
is when $F_{\alpha}=G_{\alpha}=H_{\alpha}$ 
and $P(f,\alpha,-1/2)= 2 \, P(f,\alpha,1/2)$. 
It is easy to verify that this happens for the 
$d$ quark distribution in a nonrelativistic model of the
proton with an $SU(6)$ wavefunction.  

Apart from the two particular cases illustrated above, 
Soffer's inequality
should not be expected to be saturated, and indeed it is
satisfied but not saturated in 
the model we present here.  

The model of the nucleon that we use to compute 
all ingredients appearing in 
eq.~(\ref{ingredients}) and then to evaluate 
the quark distribution functions 
is the chiral chromodielectric 
model (CCDM) \cite{Pirner}. The Lagrangian of the CCDM
reads
\begin{eqnarray}
{\cal L} &=& i\bar \psi \gamma^{\mu}\partial_{\mu} \psi
       +{g\over \chi} \, \bar \psi\left(\sigma
+i\gamma_5 \mbox{\bm $\tau \cdot \pi$} \right) \psi        \nonumber
 \\
         &+&{1\over 2}{\left(\partial_\mu\chi\right)}^2
                       -{1\over 2} M^2\chi^2
                       +{1\over 2}{\left(\partial_\mu\sigma \right)}^2
             +{1\over 2}{\left(\partial_\mu \mbox{\bm $\pi$}\right)}^2
                       -U\left(\sigma ,\mbox{\bm $\pi$} \right)   \, ,
\label{eq:in1}
\end{eqnarray}
where $U(\sigma ,\mbox{\bm $\pi$})$ is the usual mexican-hat potential,
see {\it e.g.} \cite{NF93}.
${\cal L}$  describes a system of interacting quarks, pions, sigmas and
a scalar-isoscalar chiral singlet field $\chi$.
The parameters of the model are: the chiral meson masses
$m_\pi=0.14$ GeV, $m_\sigma=1.2$ GeV, the pion decay constant
$f_\pi=93$ MeV, the quark--meson 
coupling constant $g$, and the mass $M$ of the $\chi$
field. 
The parameters $g$ and $M$, which are the only free parameters of
the model, 
 have been univoquely fixed by 
reproducing the average nucleon-delta mass and the isoscalar radius
of the proton.

The CCDM Lagrangian (\ref{eq:in1}) contains a 
single--minimum potential for the chromodielectric field 
$\chi$: $V(\chi) = \frac{1}{2} M^2 \chi^2$. A double--minimum
version of the CCDM is also widely studied and used (see for instance 
\cite{noipl}). We have checked that the structure functions 
computed in the two versions of the CCDM do not differ 
sensibly\footnote{The single--minimum CCDM seems to be 
preferable in the light of quark matter calculations \cite{qm}.}.
 
The technique used to compute the physical nucleon state appearing
in eq.~(\ref{defh1}) is based on a
double projection of the mean-field solution
on linear and angular momentum eigenstates. 
This technique was already
used to compute the static properties of the nucleon \cite{NF93}, 
the unpolarized and the longitudinally polarized
distribution functions \cite{noipl} and the nucleon
electromagnetic form factors \cite{ff}. We refer the reader to these 
references for more details. 

The intermediate states labelled by the quantum numbers $\alpha$
in eq.~(\ref{ingredients}) are also computed within the CCDM. Notice 
that they are admitted in the model, since this has no color. 
The lightest states contributing to the quark distributions
are the diquark states (scalar $ud$ and vectorial $uu, ud$). These
correspond to diagrams in which a quark is extracted and 
probed by the photon. It turns out that more massive states
($3q 1\bar q$ states arising from an antiquark insertion) 
give smaller contributions to the structure functions and 
are important only at small $x$. The antiquark distribution 
receives contributions from the $4q$ states, which correspond 
to diagrams 
with a quark insertion. We explicitly found that all these 
terms saturate with a $\sim 4 \%$ accuracy the normalization of the 
$u$ and $d$ valence distributions. This fulfillment of the 
valence number sum rule is of course a crucial check of 
the reliability of our calculation. The momentum sum rule
is satisfied as well: in our model \cite{noi,noipl}, 
at $Q_0^2$ the valence
carries about $75 \%$ of the energy-momentum, the remaining part being
carried by the mesons and the dielectric field (which, in the 
spirit of the CCDM, embodies nonperturbative glue). 
  
The transverse polarization  
distributions of quarks and antiquarks obtained
from eq.~(\ref{ingredients}) using the chiral chromodielectric model
are shown in Figs.~1-2.
We should recall that 
the distributions computed in a quark model have no dependence on the 
momentum transfer. They represent a picture
of the nucleon at some low scale $Q_0^2$, the ``model scale'' . 
Since at such low scales higher twist effects are important,
the structure functions obtained in quark models 
do not necessarily 
describe the {\it physical} nucleon at $Q_0^2$, but can be used 
as initial 
conditions for the Altarelli--Parisi evolution from 
$Q_0^2$ to a larger scale, where higher twist 
contributions are absent. In previous works \cite{noi,noipl} 
we showed how to determine 
the model scale by comparing the model prediction for the 
valence momentum with the experimental value and found 
for the CCDM  $Q_0^2 = 0.16$ GeV$^2$. 
We start from this scale the QCD evolution of our transverse
polarization densities. 

Being chirally odd, $h_1(x,Q^2)$ does not mix with gluon 
distributions, which are chirally even. Thus its $Q^2$
evolution at leading order is governed only by the 
process of gluon emission. The Altarelli--Parisi equation 
for the QCD evolution of $h_1(x,Q^2)$ 
is
\be
\frac{dh_1^{q,\bar q}(x,Q^2)}{d \log {Q^2}} = 
\frac{{\alpha}_s(Q^2)}{2 \pi} \, 
\int_x^1 \frac{dy}{y} \, P_h(y) \, h_1^{q,\bar q}(\frac{x}{y}, Q^2)\,,
\label{qcd1}
\ee
where the leading order splitting function $P_h(y)$ has been computed
by Artru and Mekhfi \cite{AM} and reads
\be
P_h(y) = \frac{4}{3} \, \left [ \frac{2}{(1+y)_{+}} - 2 + 
\frac{3}{2} \, \delta (y-1) \right ]\,.
\label{qcd2}
\ee
The Mellin transforms of the splitting function $P_h(y)$ are
the anomalous dimensions $\gamma_h^{(n)}$ which govern the 
$Q^2$ dependence of the moments of $h_1$, $h_1^{(n)}(Q^2) \equiv 
\int_0^1 \, dx \, x^{n-1} \, h_1(x,Q^2)$, according to the 
multiplicative rule
\be
h_1^{(n)}(Q^2) = h_1^{(n)}(Q^2) \, \left [ 
\frac{\alpha_s(Q_0^2)}{\alpha_s(Q^2)} 
\right ]^{\frac{6 \gamma_h^{(n)}}{33-2 n_f}}\,,
\label{qcd4} 
\ee
where $n_f$ is the number of flavors. 
In particular, since $\gamma_h^{(1)} = -2/3$, 
the first moment of $h_1$ 
and the tensor charge $\delta q \equiv \int dx \, 
(h_1^q - h_1^{\bar q})$ decrease with $Q^2$ as
\be
\delta q(Q^2) = \delta q(Q_0^2) \, 
\left [ \frac{\alpha_s(Q_0^2)}{\alpha_s(Q^2)} 
\right ]^{-4/27}\,.
\ee
Hence the $Q^2$ evolution of $h_1^{q}(x,Q^2)$ and $\delta q(Q^2)$ 
is different from that of the helicity distributions 
$\Delta q (x,Q^2)$ and of the singlet axial charge $\Delta \Sigma (Q^2)$. 
The latter is constant in $Q^2$, being related to the matrix element
of a conserved current. 
Therefore, although all existing model calculations
(including ours) give results for $h_1^q$ very close to 
those obtained for $\Delta q$, one should keep in mind 
that this scenario is valid only at the model scale $Q_0^2$. 
At typical experimental scales $h_1^q$ and $\Delta q$ are 
different in magnitude and shape as they have evolved differently
from similar inputs, and the assumption $h_1^q \simeq \Delta q$
is no longer tenable. 

The evolved distribution functions at $Q^2= 25$ GeV$^2$ 
are also shown in Figs.~1--2.
The tensor charges at this scale are: $\delta u = 0.969, \, 
\delta d = - 0.250$. 

To illustrate the different evolution of the longitudinal and the
transverse polarization distributions we compare $h_1^u$ and $\Delta u$ 
 in Fig.~3.
It is evident that,
although at $Q_0^2$ the two distributions are
almost identical, after the evolution they are largely different
at small $x$. In particular, the transverse distribution is
considerably smaller than the longitudinal one for $x < 0.1$.
A similar situation occurs for the $d$ distributions. 

Let us turn now to the possible determination of $h_1$. 
The most promising way to detect the transverse polarization 
 distribution is to measure the 
double-spin asymmetry in the Drell-Yan process with two 
transversely polarized proton beams. This quantity is given by 
(see {\it e.g.} \cite{Ji}):
\begin{equation}
A_{TT}=a_{TT}\, 
\frac{\sum_q e_q^2 h_1^q(x_a, M^2) h_1^{\bar q}(x_b,M^2) + 
( a \leftrightarrow b)}{\sum_q e_q^2 q(x_a,M^2) \bar q(x_b,M^2) 
+ ( a \leftrightarrow b) }\, ,
\label{ATT}
\end{equation}
where we have labeled by $a,b$ the two incoming protons, 
the virtuality $M^2$ of the quark and antiquark distributions
is the squared mass of the produced dilepton pair, and $x_a,x_b,M^2$ 
are related to the center of mass energy $\sqrt{s}$ by $x_a x_b = 
M^2/ s$. 
The partonic asymmetry $a_{TT}$ is calculable in 
perturbative QCD \cite{RS} and varies between $-1$ and $1$.
The double longitudinal asymmetry $A_{LL}$ has 
an expression similar to (\ref{ATT}), 
with 
the transverse distributions replaced by the 
longitudinal distributions $\Delta q (x, M^2)$.

In Fig.~4 we show our predictions for $A_{TT}/a_{TT}$ at 
$\sqrt{s} = 100$ GeV$^2$ and for various
$M^2$ values. For comparison $A_{LL}/a_{LL}$ is also shown.
Notice that the transverse asymmetry is an increasing function of the
dilepton squared mass;  however it remains about one order of magnitude
smaller than the longitudinal asymmetry. 
In Fig.~5 we present $A_{TT}$ for $x_a-x_b=0$ as a function 
of the center of mass energy: one can see that increasing 
$\sqrt{s}$ leads to a further depletion of $A_{TT}$. 
The difference between $A_{LL}$ and $A_{TT}$ 
is an effect of the different evolution of 
$h_1$ and $\Delta q$ in the small-$x$ region, which 
dominates the Drell--Yan asymmetries. 

The present calculation leads us to conclude that 
$A_{TT}$ is much smaller than it was expected on 
the basis of naive estimates. 
This is confirmed by a model--independent
study of Drell--Yan asymmetries which will be reported
in a separate paper. 
The extraction of $h_1$ is then a major challenge for 
experimentalists but is certainly worth attempting 
as it can add an important piece of information to 
our knowledge of the proton.

\vspace{1cm}
One of us (VB) would like to thank the Institute of Nuclear Theory
at the University of Washington for its hospitality and the U.S.
Department of Energy for partial support during the completion 
of this work.

\pagebreak

\pagebreak

\begin{center}
{\Large \bf Figure Captions}

\end{center}

\begin{itemize}

\item[Fig.~1]
The transverse polarization distribution of quarks 
$h_1(x)$ at the model scale
$Q_0^2=0.16$ GeV$^2$ (dashed line: $h_1^u$; dotted line: $h_1^d$) 
and 
at $Q^2=25$ GeV$^2$ (solid line: $h_1^u$;
dot-dashed line: $h_1^d$).

\item[Fig.~2]
Same as Fig. 1 for the antiquark distributions $h_1^{\bar q}$.

\item[Fig.~3]
Comparison of the evolution of the transverse polarization 
distribution
$h_1^u$ (dashed line: $Q^2= Q_0^2= 0.16$ GeV$^2$; solid line: 
$Q^2=25$ GeV$^2$)
and of the longitudinal polarization distribution  $\Delta u$ 
(dotted line: $Q^2 = Q_0^2= 0.16$ GeV$^2$; dot-dashed line: 
$Q^2=25$ GeV$^2$).

\item[Fig.~4]
Predictions for the Drell-Yan double 
transverse asymmetry
$A_{TT}/a_{TT}$ (dot-dashed line: $M^2=50$ GeV$^2$; dashed line: 
$M^2=25$ GeV$^2$; solid line: $M^2=10$ GeV$^2$). For comparison, 
the double
longitudinal asymmetry  $A_{LL}/a_{LL}$ is shown 
for $M^2= 10$ GeV$^2$ (dotted line). All curves are obtained with
$\sqrt{s}=100$ GeV.

\item[Fig.~5]
Dependence on $M^2$ of the transverse double spin asymmetry at 
$x_a-x_b=0$ (dot-dashed line: $\sqrt{s}=100$ GeV; dashed line: 
$\sqrt{s}=300$ GeV; solid line: $\sqrt{s}=500$ GeV).

\end{itemize}

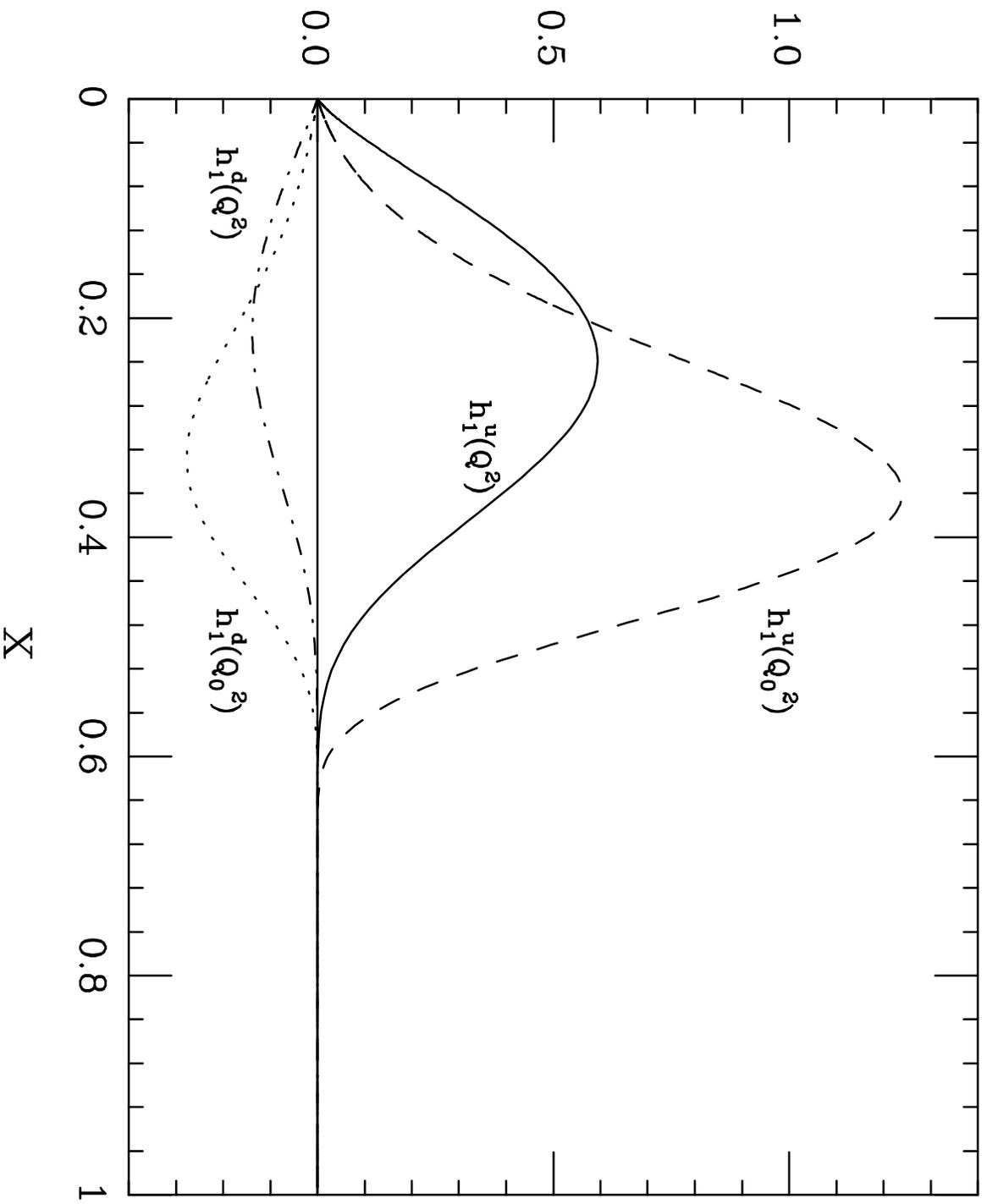
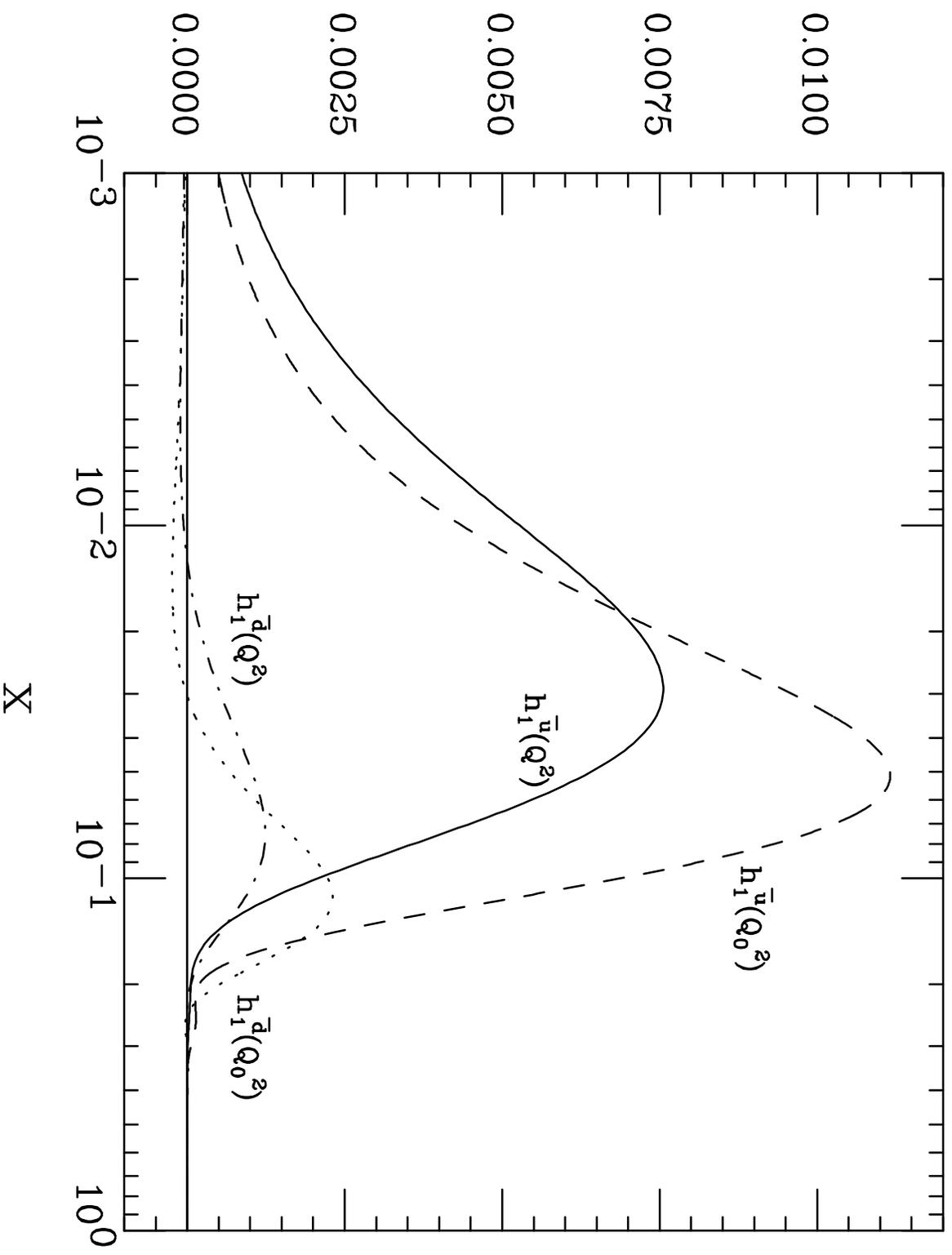
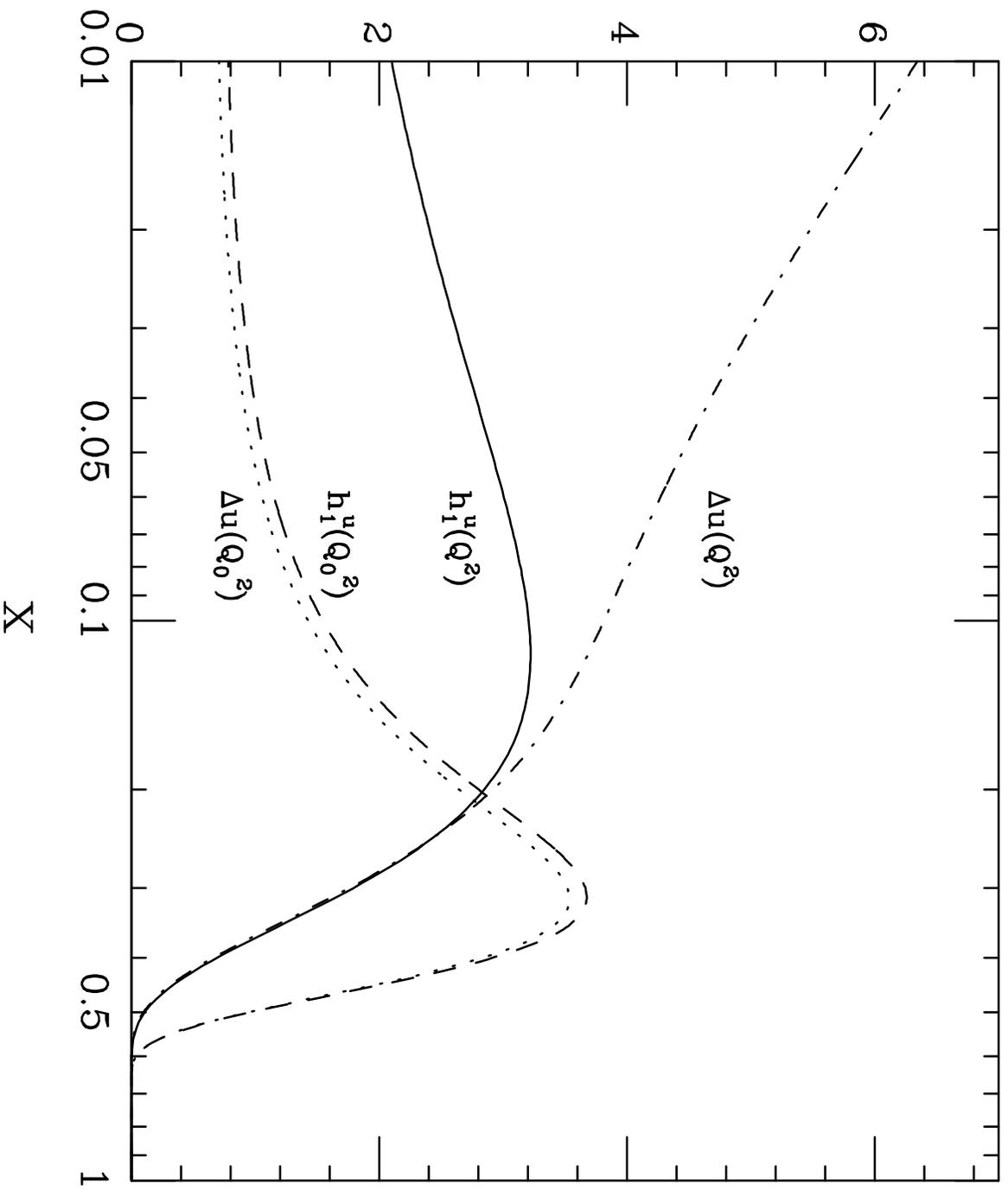
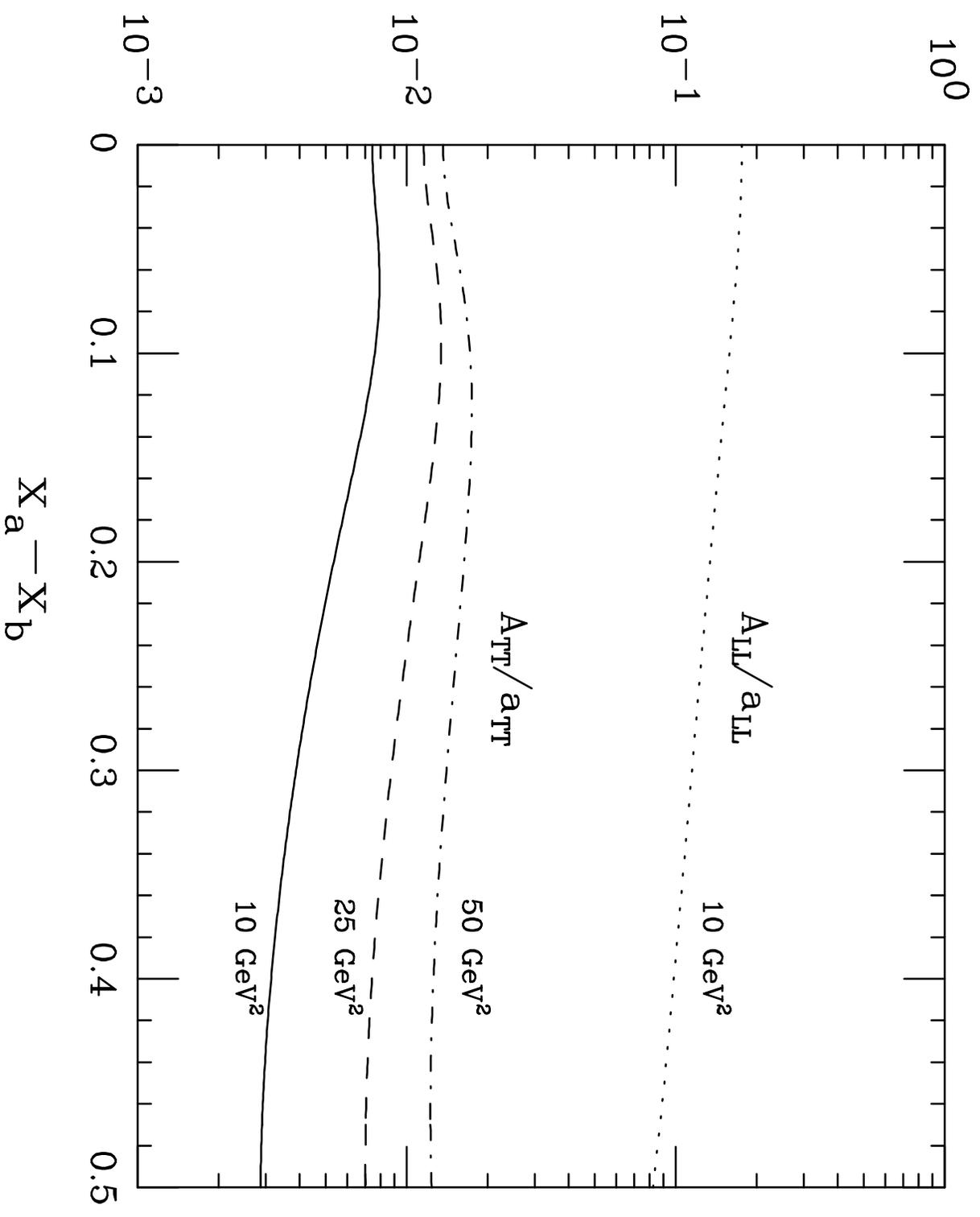
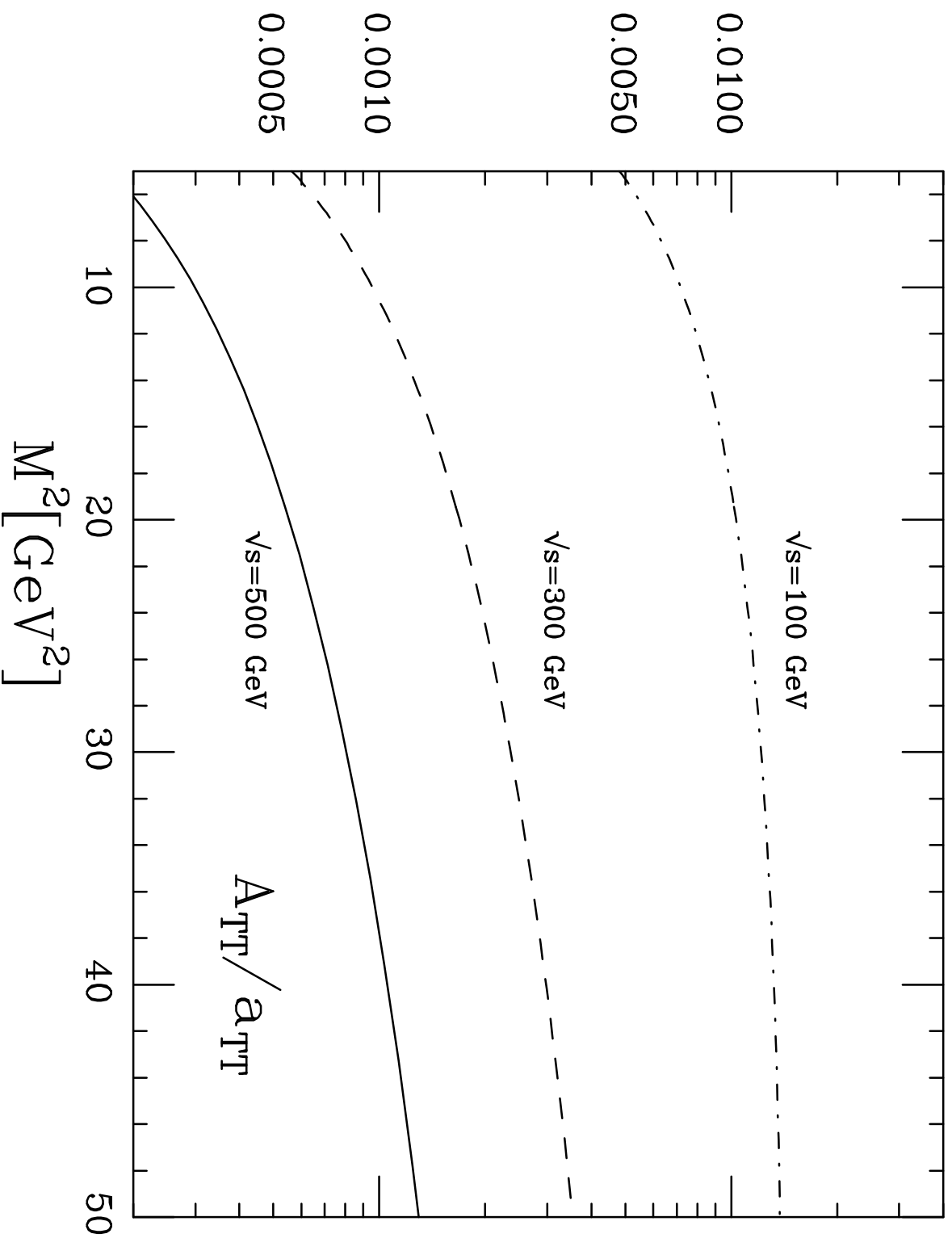
\end{document}